\documentclass[10pt, conference]{IEEEtran}

\ifCLASSOPTIONcompsoc
  \usepackage[nocompress]{cite}
\else
  \usepackage{cite}
\fi

\usepackage{balance}
\usepackage{array,enumerate}
\usepackage{multirow}
\usepackage{graphicx}
\usepackage{algorithm}
\usepackage{algorithmic}
\usepackage{amssymb}
\usepackage{amsmath}
\usepackage{url}
\usepackage{threeparttable}
\usepackage{scrextend}
\usepackage{bm}
\usepackage{tikz}

\newcommand{\subcaption}[1]{\centerline{{\scriptsize
  #1}}\vspace{10pt}}
\newlength{\minipagewidth}
\newlength{\figurewidthFour}

\begin{document}

\title{Towards Distributed Machine Learning in Shared Clusters: A Dynamically-Partitioned Approach}


\author{
    \IEEEauthorblockN{Peng\ Sun\IEEEauthorrefmark{1},
    Yonggang\ Wen\IEEEauthorrefmark{1},
    Ta\ Nguyen Binh Duong\IEEEauthorrefmark{1}
    and Shengen\ Yan\IEEEauthorrefmark{2}
    }
    \IEEEauthorblockA{\IEEEauthorrefmark{1} Nanyang Technological University, Singapore, \IEEEauthorrefmark{2} Sensetime Group Limited}
    \IEEEauthorblockA{\{sunp0003, ygwen, donta\}@ntu.edu.sg, yanshengen@sensetime.com}
}

\maketitle

\begin{abstract}

Many cluster management systems (CMSs) have been proposed to share a single cluster with multiple distributed computing systems. However, none of the existing approaches can handle  distributed machine learning (ML) workloads given the following criteria: high resource utilization, fair resource allocation and low sharing overhead. To solve this problem, we propose a new CMS named Dorm, incorporating a dynamically-partitioned cluster management mechanism and an utilization-fairness optimizer. Specifically, Dorm uses the container-based virtualization technique to partition a cluster, runs one application per partition, and can dynamically resize each partition at application runtime for resource efficiency and fairness. Each application directly launches its tasks on the assigned partition without petitioning for resources frequently, so Dorm imposes flat sharing overhead. Extensive performance evaluations showed that Dorm could simultaneously increase the resource utilization by a factor of up to $2.32$, reduce the fairness loss by a factor of up to $1.52$, and speed up popular distributed ML applications by a factor of up to $2.72$, compared to existing approaches. Dorm's sharing overhead is less than $5\%$ in most cases.

\end{abstract} 


\section{Introduction}\label{sec:introduction}

A diverse array of distributed computing systems (DCSs) have emerged to handle  various big data applications. Prominent examples include Hadoop and Spark. To offer better performance when training  machine learning (ML) models, a lot of distributed ML systems have been proposed based on the ParameterServer (PS) framework, such as MxNet \cite{chen2015mxnet}, MPI-Caffe \cite{jia2014caffe},  TensorFlow \cite{abadi2016tensorflow} and Petuum \cite{xing2015petuum}. These systems could decompose an application into a set of small tasks and execute them on multiple nodes in parallel \cite{hu2014toward}.

Many cluster management systems (CMSs) have been proposed to run multiple DCSs in the same cluster for two reasons. First, users can  pick the best DCS for each application \cite{hindman2011mesos}. Second, cluster sharing could considerably improve the cluster resource utilization and application performance \cite{delimitrou2014quasar}. Existing CMSs can be classified into six categories based on their cluster management strategies. Specifically, Infrastructure-as-a-Service (IaaS) approaches (e.g., OpenStack \cite{sefraoui2012openstack}) can share clusters at the level of DCSs. For example, we can create a set of virtual machines (VMs) for Spark, and run all Spark applications in this virtual cluster. Monolithic, two-level, shared-state, fully-distributed and hybrid approaches can allocate cluster resources  at the level of applications and tasks, such as Yarn \cite{vavilapalli2013apache}, Mesos \cite{hindman2011mesos}, Quasar \cite{delimitrou2014quasar}, etc.

\setlength{\minipagewidth}{0.2 \textwidth}
\setlength{\figurewidthFour}{\minipagewidth}
\begin{figure}  
    \centering
    \begin{minipage}[t]{\minipagewidth}
    \begin{center}
    \includegraphics[width=\figurewidthFour]{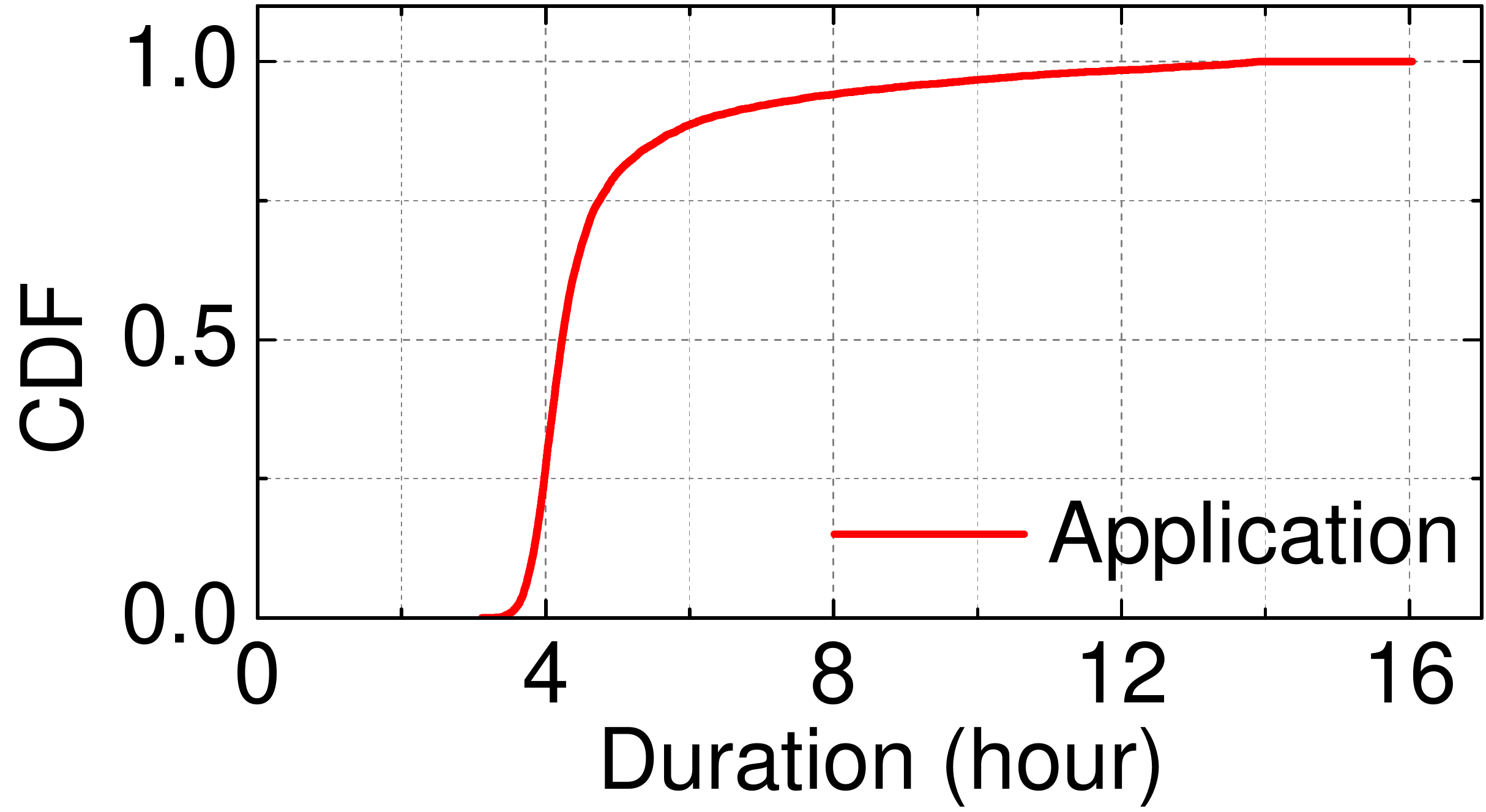}
    \end{center}
    \end{minipage}
    \centering
    \hspace{2.2 mm}
    \begin{minipage}[t]{\minipagewidth}
    \begin{center}
    \includegraphics[width=\figurewidthFour]{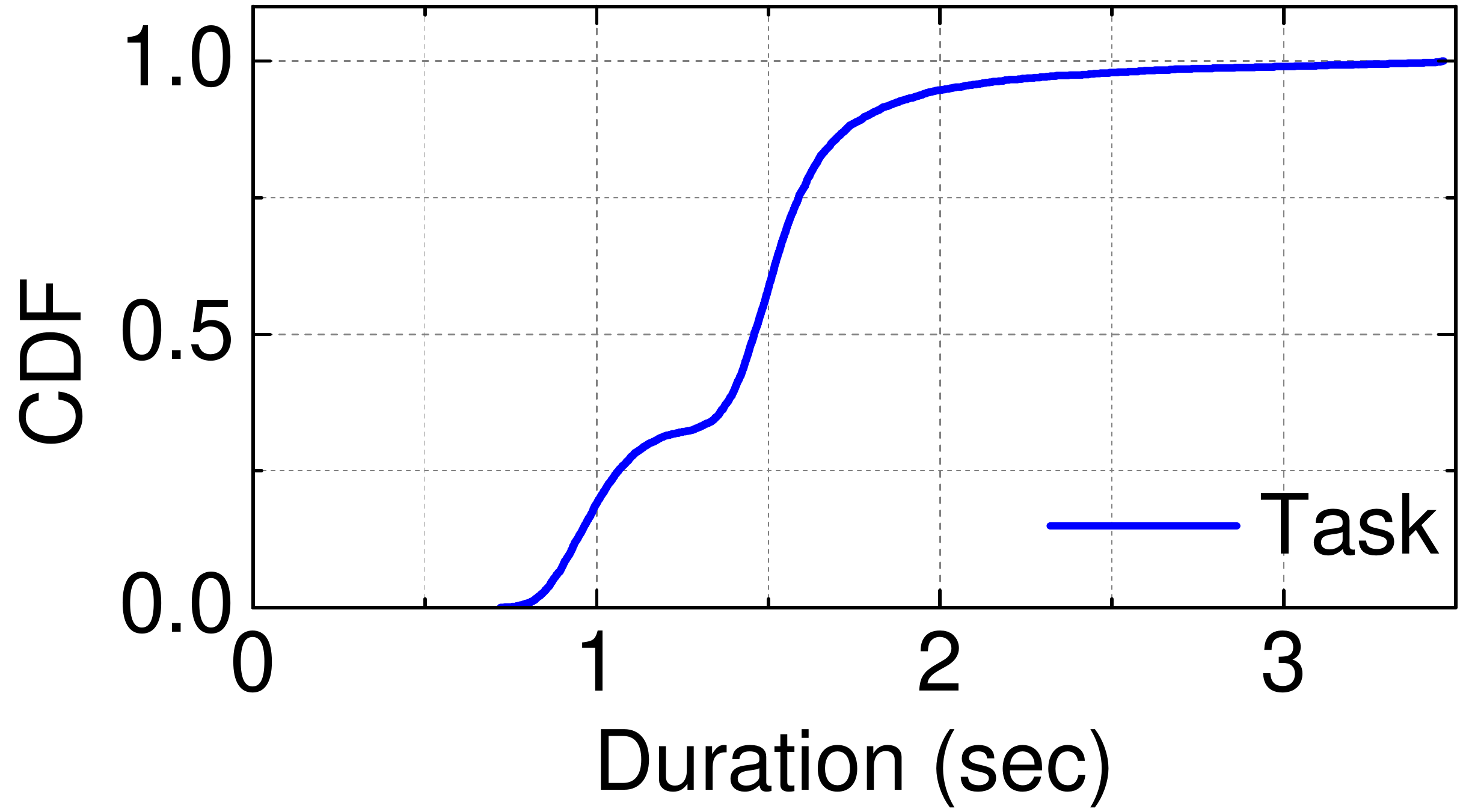}
    \end{center}
    \end{minipage}
    \centering
    \caption{CDF of distributed ML application and task duration.}
\label{Fig: ml_time}
\end{figure}

In this work, we consider the problem of running multiple and  diverse distributed ML workloads in a single cluster. 
For  example, Sensetime Group Limited operates several clusters with thousands of CPUs/GPUs, and uses various distributed ML systems to train  ML models on them.  A workload analysis from one production cluster suggests that distributed ML applications usually have long application duration and very short task duration.  Figure \ref{Fig: ml_time} shows that about $90\%$ of distributed ML applications run more than $6$ hours; and about $50\%$ of distributed ML tasks use less than $1.5$s.

However, none of the existing CMSs can efficiently handle  distributed ML workloads in a shared cluster given three criteria:  high resource utilization, low fairness loss\footnote{Low fairness loss indicates that each applications could receive a fair share of resources. Its detail definition can be found in Section IV.} and low sharing overhead\footnote{Sharing overhead denotes the percentage of an application's additional running time imposed by a CMS.}. 
IaaS CMSs cannot handle distributed ML workloads at the level of DCSs, since popular distributed ML systems do not have multi-application support. For example, we need to manually allocate the resources of a TensorFlow virtual cluster to multiple applications. Monolithic, two-level, shared-state, fully-distributed and hybrid CMSs can only statically allocate resources to distributed ML applications, and do not allow them to dynamically scale up/down or scale out/in based on the global cluster state,  resulting in low resource utilization and high fairness loss \cite{hindman2011mesos}. 


In this paper, we propose a new CMS named Dorm to handle multiple distributed ML workloads in a shared cluster with  two techniques: a dynamically-partitioned cluster management mechanism and an utilization-fairness optimizer. Dorm uses the container-based virtualization technique to partition a cluster, and runs one application per partition. Each application places its tasks on the assigned partition  without petitioning for resources, so Dorm imposes low sharing overhead. When detecting newly submitted or completed applications, Dorm could adjust existing resource allocations to consistently keep high resource utilization and low fairness loss.

We implement Dorm using Docker and Cloud3dView \cite{yin2013cloud3dview}, and integrate it with four  widely used distributed ML systems: Petuum, MxNet, TensorFlow and MPI-Caffe. 
Extensive evaluations on a working testbed showed that Dorm  could simultaneously improve the resource utilization by a factor of up to 2.32, reduce the fairness loss by a factor of up to 1.52, and speed up popular  distributed ML applications  by a factor of up to 2.72, compared to existing approaches. In most cases, Dorm could limit the  sharing  overhead within $5\%$.

\section{Background and Related Work} \label{sec: related_work}
In this section, we introduce distributed ML, review and analysis existing cluster management systems.

\subsection{Distributed ML: A Primer}


\setlength{\minipagewidth}{0.45\textwidth}
\setlength{\figurewidthFour}{\minipagewidth}
\begin{figure} 
    \centering
    \begin{minipage}[t]{\minipagewidth}
    \begin{center}
    \includegraphics[width=\figurewidthFour]{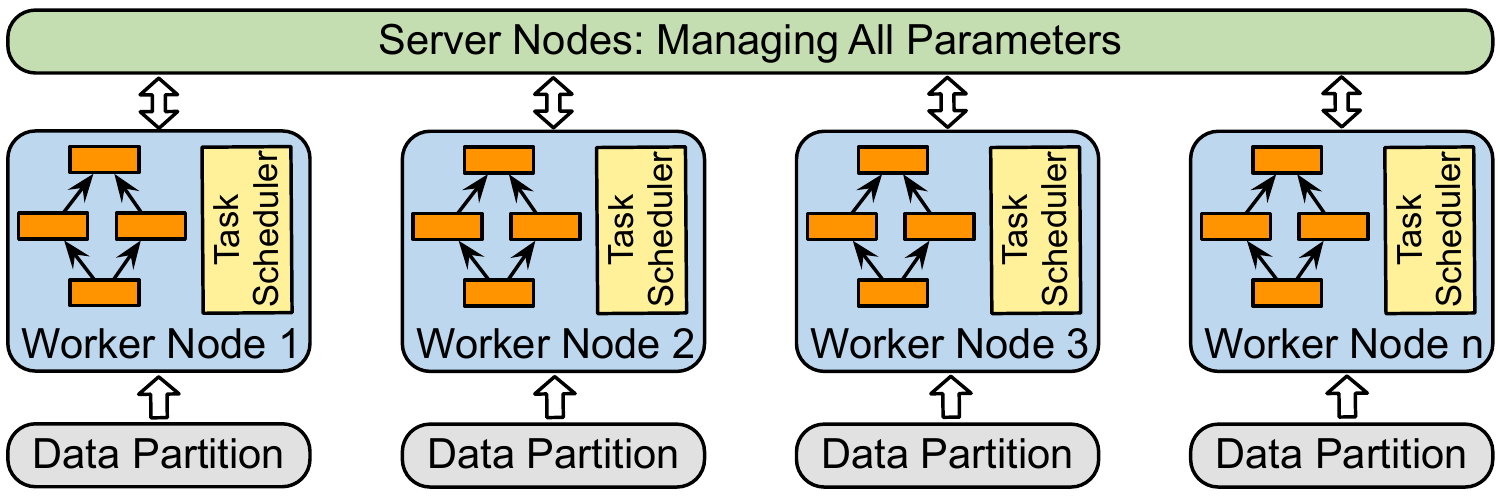}
    \end{center}
    \end{minipage}
    \centering
    \caption{The PS framework's architecture.}
\label{Fig: distributed_ML_arch_mapreduce}
\end{figure}

The goal of ML is to learn  models from training datasets, and use them to make predictions on new data. To handle big training datasets and big models, many distributed ML systems have been proposed based on the PS framework. As shown in Figure \ref{Fig: distributed_ML_arch_mapreduce}, the PS framework can scale to large cluster deployment by having {worker} nodes performing data-parallel computation, and having {server} nodes maintaining globally shared \emph{parameters} of ML models.  Each worker node contains a TaskScheduler to place tasks on the local node based on a specific policy, such as   Bulk Synchronous Parallel (BSP) or Stale Synchronous Parallel (SSP) \cite{xing2015petuum}.

\subsection{Related Work: Cluster Management Systems}

CMSs are designed to run multiple DCSs in a single cluster.
As shown in Figure \ref{Fig: existing approaches},  existing CMSs can be classified into six categories based on their cluster management strategies. These approaches could  perform resource allocation at three levels: DCS, application and task. Resource  allocation refers to determining the amount of resources offered to  applications, and selecting specific resources from  servers to satisfy user-supplied placement preferences \cite{delimitrou2014quasar}.

\setlength{\minipagewidth}{0.485\textwidth}
\setlength{\figurewidthFour}{\minipagewidth}
\begin{figure} 
    \centering
    \begin{minipage}[t]{\minipagewidth}
    \begin{center}
    \includegraphics[width=\figurewidthFour]{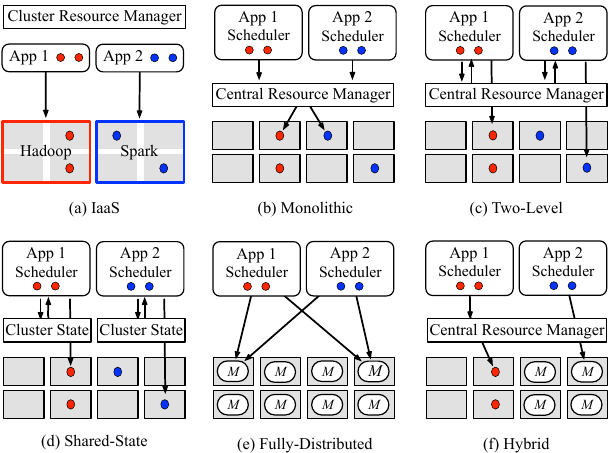}
    \end{center}
    \end{minipage}
    \centering
    \caption{Taxonomy of existing CMSs. Circles represent tasks; gray boxes represent cluster servers;  and $M$ denotes a distributed resource manager. }
\label{Fig: existing approaches}
\end{figure}

\textbf{IaaS CMSs}, such as OpenStack \cite{sefraoui2012openstack}, use VMs to partition a cluster, run one DCS per partition, and let each DCS to manage and schedule submitted applications \cite{jin2013empirical}. 

\textbf{Monolithic CMSs}, such as Yarn \cite{vavilapalli2013apache}, Quasar \cite{delimitrou2014quasar} and Borg \cite{verma2015large}, use a centralized resource manager to perform resource allocation for all applications with cluster-wide visibility.

\textbf{Two-level CMSs}, such as Mesos \cite{hindman2011mesos}, use a central cluster resource manager and application-specific schedulers to jointly perform resource allocation. The central manager gives each application a set of resource offers, and let the application-specific scheduler decide whether to accept them.

\textbf{Shared-state CMSs} let each application maintain a copy of the cluster state, and compete for resources using lock-free optimistic concurrency control, as in Omega \cite{schwarzkopf2013omega} and Apollo \cite{boutin2014apollo}. These approaches could offer high resource allocation quality without strict fairness guarantees due to the lack of centralized resource management. 

\textbf{Fully-distributed CMSs}, such as Sparrow \cite{ousterhout2013sparrow}, use many independent resource managers to serve applications' resource requests  with local, partial and stale cluster state. This approach can achieve millisecond scheduling latency per request.

\textbf{Hybrid CMSs} combine distributed resource managers with a centralized cluster scheduler, as in Hawk \cite{delgado2015hawk} and Mercury \cite{karanasos2015mercury}. Applications can  obtain strong execution guarantees from the centralized scheduler, or trade strict guarantees for millisecond scheduling latency from  distributed managers.


IaaS CMSs share cluster resources at the level of DCSs. This approach requires that DCSs could manage and schedule multiple applications. Monolithic, two-level, shared-state, fully-distributed and hybrid CMSs support both app-level and task-level resource allocation. In app-level  mode, each application would reserve all allocated resources until completion. In task-level  mode, applications would use acquired resources to run a single task, release them as soon as the task completes, and petition for new resources to launch uncompleted tasks.

\subsection{Performance Analysis}

Existing approaches cannot simultaneously achieve high resource utilization, low fairness loss and low sharing overhead when handling  distributed ML workloads.
IaaS CMSs cannot work in conjunction with popular  distributed ML systems (e.g., TensorFlow), which have no multi-application support. In app-level sharing mode, monolithic, two-level, shared-state, fully-distributed and hybrid CMSs cannot dynamically adjust existing resource allocations to consistently keep  high resource utilization and low fairness loss. 
In task-level sharing mode, monolithic and two-level CMSs impose high sharing overhead, since each task must wait until receiving suitable resources. 
For example, in a 100-node Mesos cluster, our experiments  showed that the average scheduling latency per task is about $430$ms, which represents significant sharing overhead for short distributed ML tasks. Shared-state, fully-distributed and hybrid CMSs introduce concurrency control and distributed scheduling to reduce the sharing overhead at the cost of high  fairness loss, due to the lack of centralized resource management. 

In practices, existing CMSs could only statically allocate user-specified resources to distributed ML applications, as in TensorFlow-on-Mesos and MxNet-on-Yarn. When submitting a new  application, users must manually specify its resource demands, including the number of worker nodes, and the amount of CPUs, GPUs and RAM per worker node.

\section{Dorm: A Dynamically-Partitioned Approach} \label{sec: system}

We propose a new CMS named Dorm to efficiently handle multiple and diverse distributed ML workloads in a single cluster using two techniques: a dynamically-partitioned cluster management mechanism and an utilization-fairness optimizer.  In this section, we focus on the first technique.

\subsection{System Architecture} 

Figure \ref{Fig: system_architecture} shows Dorm's system architecture.  Dorm is a type of the monolithic CMS, which  contains a central DormMaster and a set of DormSlaves.

\subsubsection{\textbf{DormMaster}}

The  {DormMaster} centrally manages all cluster resources, and exposes them to  applications. 
It uses \emph{containers}\footnote{The \emph{container} is a logical bundle of  resources on a server, for example $\left\langle \text{2 CPUs, 1 GPU, 8GB RAM} \right\rangle$.} to  partition a cluster, and gives each  application a partition. 
The utilization-fairness optimizer is a  module of the DormMaster to make resource allocation decisions. 

\subsubsection{\textbf{DormSlave}}

The {DormSlave} manages local resources of a cluster server. It  reports the amount of available resources of a cluster server to the {DormMaster}, and uses \emph{containers} to share a cluster server with multiple applications.

\subsubsection{\textbf{Application}}

Dorm is designed to host distributed ML applications.
Since modern distributed ML systems usually use distributed scheduling mechanisms as shown in Section II, Dorm deploys a {TaskExecutor} and a {TaskScheduler} on each \emph{container}.
The {TaskExecutor}  is the basic unit to execute tasks.
The {TaskScheduler} is charge of placing tasks of an application  on the local {TaskExecutor}.  

\subsubsection{\textbf{Container}}

\emph{Containers} of the same application would have uniform, constant resource demands for two reasons. First,  distributed ML applications could balance the workloads across all TaskExecutors by equally partitioning the training datasets.  Second, distributed ML applications usually use iterative methods to train models without changing resource demands during application runtime.

\setlength{\minipagewidth}{0.45\textwidth}
\setlength{\figurewidthFour}{\minipagewidth}
\begin{figure} 
    \centering
    \begin{minipage}[t]{\minipagewidth}
    \begin{center}
    \includegraphics[width=\figurewidthFour]{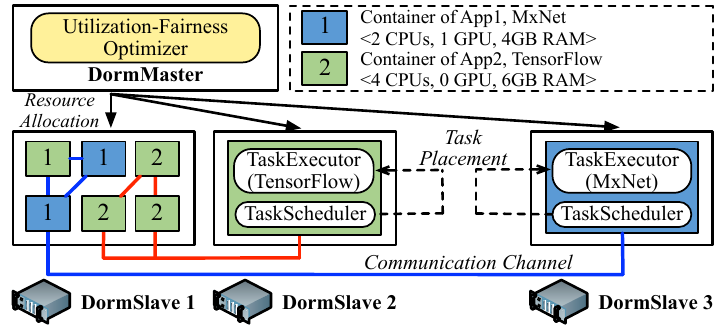}
    \end{center}
    \end{minipage}
    \centering
    \caption{Dorm's system  architecture. In this example, a MxNet-based application and a TensorFlow-based application share a cluster with 3 servers.}
\label{Fig: system_architecture}
\end{figure}

\subsection{{Application Submission}}

To submit a new distributed ML application to Dorm, users need to provide a 6-tuple as follows:
$$(executor, \bm{d}, w, n_{max}, n_{min}, cmd),$$
where $executor$ is a string (e.g., ``MxNet'') to indicate the required computation engine; $\bm{d}$ is the resource demand vector (e.g., $\left\langle \text{2 CPUs, 1 GPU, 8GB RAM} \right\rangle$) per \emph{container}; $w$ is an integer to show this application's weight; $n_{max}$ and $n_{min}$ represent the maximum and minimum  numbers of \emph{containers} this application could have;  $cmd$ specifies the scripts used to start and resume this  application.

\subsection{Dynamically-Partitioned Resource Management}  \label{sec: DPC}

Dorm performs resource allocation in a dynamic manner at the level of applications. In a nutshell, it gives each application a partitioned cluster, and can dynamically resize each partition.

\subsubsection{\textbf{Making Resource Allocation Decisions}}

When detecting newly submitted or completed applications, 
the utilization-fairness optimizer
determines new resource allocations for resource efficiency and fairness based on the algorithm detailed in Section IV.

\subsubsection{\textbf{Adjusting Existing Resource Allocations}}

Dorm could enforce new resource allocations by adjusting existing  ones:  creating and destroying \emph{containers} on particular servers. However, popular distributed ML applications cannot automatically take advantage of newly acquired resources, or keep running with revoked resources. To address this problem, we propose a checkpoint-based resource adjustment protocol.
Specifically, when adjusting  an application's resources, Dorm would firstly save its state to a reliable  storage system (e.g., the Lustre file system).  
Then, Dorm kills this application, and creates/destroys \emph{containers} on corresponding servers. 
Finally, Dorm resumes the killed application from the saved state with new resource allocations. In this way, distributed ML applications can dynamically scale up or down without recomputing from the first iteration.
 
\setlength{\minipagewidth}{0.45\textwidth}
\setlength{\figurewidthFour}{\minipagewidth}
\begin{figure} 
    \centering
    \begin{minipage}[t]{\minipagewidth}
    \begin{center}
    \includegraphics[width=\figurewidthFour]{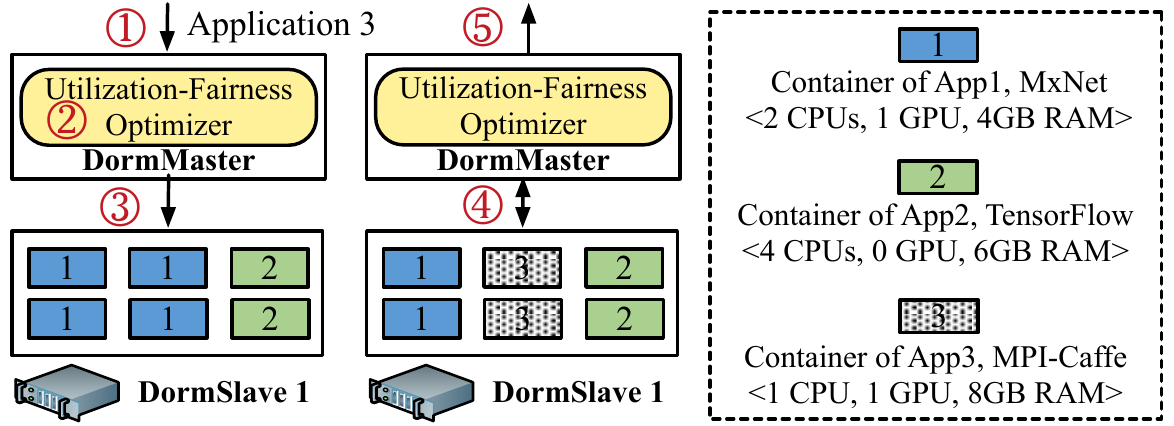}
    \end{center}
    \end{minipage}
    \centering
    \caption{An example to show how Dorm allocates resources to  applications.}
\label{Fig: system_architecture_2}
\end{figure}

\subsubsection{\textbf{An Example}}

Figure \ref{Fig: system_architecture_2} shows an example of how Dorm allocates resources to applications. In step (1), an user submits a new application to Dorm with following information: 
\begin{equation} \nonumber
\begin{split}
& executor = \text{``MPI-Caffe''}, \bm{d}  =  \left\langle \text{1 CPU, 1 GPU, 8GB RAM} \right\rangle, \\
  & w = 2, n_{max} = 5, n_{min}  = 1, cmd = \text{[``start.sh'', ``resume.sh'']}.
\end{split}
\end{equation}
In step (2), the utilization-fairness optimizer determines that all applications should have 2 \emph{containers} on DormSlave 1. In step (3), the {DormMaster} enforces new resource allocations by destroying $2$ \emph{containers} of App2 and creating $2$ \emph{containers} for APP3 on DormSlave 1. In this step, Dorm saves App2's state to a reliable storage system and kill it. In step (4), the {DormMaster} configures TaskExecutors and TaskSchedulers on new \emph{containers}, starts APP3, and resumes  APP2.  In step (5), the {DormMaster} returns  APP3's status to the user.

\subsection{Task Placement} 

Dorm uses application-specific schedulers to place applications' tasks on assigned partitions.
Since modern distributed ML systems use distributed scheduling mechanisms, Dorm deploys a TaskScheduler and a TaskExecutor per \emph{container}. During application runtime, each  TaskScheduler is in charge of placing tasks of an application on the local TaskExecutor. 
Therefore, application-specific schedulers would not request for resources to launch individual tasks, leading to  low scheduling latency and low sharing overhead.

\section{Utilization-Fairness Optimizer} \label{sec: optimizer}

In this section, we show how the utilization-fairness optimizer makes  resource allocation decisions. Table \ref{Table: symbol} shows used symbols and their definitions in this section.

\subsection{Objectives}

We consider a cluster with $m$ types of hardware resources. When allocating cluster resources to the running application set  $\mathcal{A}^t$ at time $t$, we aim to achieve high resource utilization and low fairness loss with low resource adjustment overhead.   

\subsubsection{\textbf{Resource Utilization}}

The cluster's resource utilization is defined as the sum of all $m$ types of hardware resources' utilization, which can be represented as follows:
\begin{equation} 
\begin{split}
\text{ResourceUtilization}(t) = \sum\nolimits_{k \in \mathcal{M}}{u_k^t},
\end{split}
\end{equation}
where $u_k^t = \sum\nolimits_{i \in \mathcal{A}^t} \sum\nolimits_{j \in \mathcal{B}} \frac { x_{i,j}^t d_{i,k}}{\sum_{h \in \mathcal{B}}c_{h,k}}$ denotes resource $k$'s utilization at time $t$.

\renewcommand\arraystretch{1}
\begin{table} 
\centering
\caption{Summary of notations used.}
\label{Table: symbol}
\resizebox{0.49\textwidth}{!}{
\begin{tabular}{|p{0.8cm} @{} | p{8cm} | }
\hline 
$x_{i, j}^t$    & application $i$'s \emph{container} number on DormSlave $j$ at time $t$.  \\
$l_{i}^t$       & application $i$'s fairness loss at time $t$.  \\
$r_{i}^t$       & application $i$'s resources are adjusted at time $t$.  \\
${u}_i^t$                    & resource $i$'s utilization at time $t$\\
$n_{i}^{max}$   & application $i$'s maximum  \emph{container} number  \\ 
$n_{i}^{min}$   & application $i$'s minimum  \emph{container} number  \\ 
$\theta_1$         & threshold of fairness loss \\
$\theta_2$         & threshold of resource adjustment overhead  \\
${\hat s}^t_i$    &  application $i$'s theoretical resource share based on DRF at time $t$ \\
${s}^t_i$    & application $i$'s   actual resource share at time $t$ \\
$d_{i,j}$         & application $i$'s resource demand on resource $j$ \\
$c_{i,j}$         & DormSlaves $i$'s resource capacity on resource $j$ \\ 
$\mathcal{B}$  & set of DormSlaves, $\mathcal{B} = \{1, 2, \ldots, b\}$  \\
$\mathcal{M}$  & set of resource types, $\mathcal{M} = \{1, 2, \ldots, m\}$\\
$\mathcal{A}^{t}$ & set of applications running at time $t$ \\
\hline
\end{tabular}
}
\end{table}

\subsubsection{\textbf{Fairness Loss}}

Fairness indicates that each application could receive a fair share of resources based on a particular fairness policy.
In this paper, we use  dominant resource fairness (DRF) \cite{ghodsi2011dominant} as the fairness policy. DRF seeks to maximize the minimum dominant share\footnote{Dominant resource is the mostly heavily demanded resource required by an application, and dominant share is the share of the dominant resource.} across all  applications. 
Let  ${\hat s}^t_i$ denote application $i$'s theoretical dominant  share derived from DRF based on the algorithms proposed in \cite{ghodsi2011dominant}. Let ${s}^t_i$ denote application $i$'s actual dominant share.  The cluster's fairness loss is defined as the sum of all applications' fairness loss, which can be represented as follows:
\begin{equation} 
\begin{split}
\text{FairnessLoss}(t) 
=  \sum\nolimits_{i \in \mathcal{A}^t}{l_i}  = \sum\nolimits_{i \in \mathcal{A}^t}{\left | s_i^t - \hat s_i^t \right|},
\end{split}
\end{equation}
where  $s_i^t = \underset{k \in \mathcal{M}} \max (\frac{d_{i,k} \sum\nolimits_{j \in \mathcal{B}}  x_{i,j}^t} {\sum_{h \in \mathcal{B}} c_{h,k}})$.

\subsubsection{\textbf{Resource Adjustment Overhead}}
The cluster's resource adjustment overhead is measured by the number of affected applications, which would be killed and resumed, to enforce the newly computed resource allocations.  
Let $r_i^t$ denote whether Dorm would adjust application $i$'s resources:
\begin{equation} \label{Eq: def_r} 
r_i^t = 
\begin{cases}
0, & \text{if} \; x_{i,j}^{t-1} = x_{i,j}^t, \; \forall j \in \mathcal{B} \\ 
1, & \text{if} \; x_{i,j}^{t-1} \neq  x_{i,j}^t, \; \exists j \in \mathcal{B} 
\end{cases}
.
\end{equation}
If $r_i^t = 0$, Dorm would not create or destroy \emph{containers} on any cluster servers for  application $i$, and  vice versa. It should be noted that the newly launched  and completed applications at time $t$ would not be considered as the affected applications due to resource adjustment. Therefore, the cluster's resource adjustment overhead  can be represented as follows:
\begin{equation} \label{Eq: def_adjust}
\text{ResourceAdjustmentOverhead}(t) = \sum\nolimits_{i \in \mathcal{A}^t \cap \mathcal{A}^{t-1}} {r_i^t},
\end{equation}
where $\mathcal{A}^t \cap \mathcal{A}^{t-1}$ is the set of applications running at both time $t-1$ and $t$. 

\subsection{Problem Formulation}

Dorm  determines the number of \emph{containers} offered to  applications, and the location of each \emph{container}.  
We formulate this  problem  as a multi-objective optimization problem as follows:
\begin{alignat}{3}
\textbf{P1:} \; \max \;
& \bigg [ {\sum\nolimits_{i \in \mathcal{M}} u_{i}},  -{\sum\nolimits_{i \in \mathcal{A}^t}{l_i}}, && -{\sum\nolimits_{i \in  \mathcal{A}^t }{r_i}} \bigg ]    \label{Eq: objective_all}  \\ 
\text{s.t.}  \; 
& \sum\nolimits_{i \in \mathcal{A}^t} x_{i,j}^t d_{i,k} \leq c_{j,k}, 
&& \forall k \in \mathcal{M},  \forall j \in \mathcal{B} \label{Eq: resource_capacity}
\\
& \sum\nolimits_{j \in \mathcal{B}} x_{i,j}^t \leq n_{i}^{max}, 
&& \forall i \in \mathcal{A}^t   \label{Eq: uppernumber}
\\
& \sum\nolimits_{j \in \mathcal{B}}  x_{i,j}^t \geq n_{i}^{min}, 
&&  \forall i \in \mathcal{A}^t  \label{Eq: lowernumber}
\\
& x_{i,j}^t \in \mathcal{Z}_0^+, 
&&  \forall i \in \mathcal{A}^t,
\forall j \in \mathcal{B} \label{Eq: integer}
\end{alignat}
Equation \ref{Eq: objective_all} is the objective function, which shows that we want to maximize resource utilization, minimize fairness loss and minimize resource adjustment overhead. We have several constraints.  Equation \ref{Eq: resource_capacity} indicates that each cluster server cannot exceed  its resource capacity. Equation \ref{Eq: uppernumber} and \ref{Eq: lowernumber} constraint  the maximum and minimum numbers of \emph{containers} an application can have. 
Equation \ref{Eq: integer} shows that $x_{i,j}^t$ is an integer variable.

We then transform \textbf{P1} into a  MILP problem as follows:
\begin{alignat}{3}
\textbf{P2:} \; \max \;
& \sum\nolimits_{k \in \mathcal{M}} \sum\nolimits_{i \in \mathcal{A}^t} \sum\nolimits_{j \in \mathcal{B}}  \frac {x_{i, j}^t d_{i,k}}{\sum_{h \in \mathcal{B}}c_{h,k}} \label{Eq: final_objective}
\\
\text{s.t.}  \;
& l_i^t \geq s_i^t - \hat s_i^t,  
\quad \; \; \, \quad \qquad \quad \; \; \;
\forall i \in \mathcal{A}^t   \label{Eq: fairness_loss_1}
\\
& l_i^t \geq {\hat s}_i^t - s_i^t, 
\quad \; \; \, \quad \qquad \quad \; \; \;
\forall i \in \mathcal{A}^t   \label{Eq: fairness_loss_2}
\\
& Mr_i^t \geq x_{i,j}^{t-1} - x_{i,j}^t,  
 \forall j \in \mathcal{B}, \forall i \in \mathcal{A}^t \cap \mathcal{A}^{t-1} \label{Eq: adjust_1}
\\
& Mr_i^t \geq x_{i,j}^t - x_{i,j}^{t-1},
 \forall j \in \mathcal{B}, \forall i \in \mathcal{A}^t \cap \mathcal{A}^{t-1}  \label{Eq: adjust_2}
\\
& \sum\nolimits_{i \in \mathcal{A}_{r}^t} {l_i^t}  \leq  \big \lceil \theta_1 \times 2 m  \big \rceil,
\label{Eq: threshold_l}
\\
& \sum\nolimits_{i \in \mathcal{A}_{r}^t} {r_i^t} \leq \lceil \theta_2 \times |\mathcal{A}^{t} \cap \mathcal{A}^{t-1}| \rceil, 
\label{Eq: threshold_r}
\\
& l_{i}^t \in \mathcal{R}_0^+, 
\quad \quad \quad \quad \quad \; \; \;  \; \; \, \quad \; 
\forall i \in \mathcal{A}^t \label{Eq: range_l}
\\
&r_{i}^t \in \{0, 1\}, 
\quad \quad \quad \quad \quad \;  \, \quad \; 
\forall i \in \mathcal{A}^t \label{Eq: range_r}
\\
& (\ref{Eq: resource_capacity}),  (\ref{Eq: uppernumber}), (\ref{Eq: lowernumber}), (\ref{Eq: integer}). \nonumber
\end{alignat}
In this formulation, we choose resource utilization as the objective to be maximized; fairness loss and resource adjustment overhead are constrained to be no greater than some given thresholds. 
Equation \ref{Eq: fairness_loss_1} and \ref{Eq: fairness_loss_2} are used to linearize  $l_i^t$. Equation \ref{Eq: adjust_1} and \ref{Eq: adjust_2} are used to linearize $r_i^t$ with a big number $M$.  Equation \ref{Eq: threshold_l} and \ref{Eq: threshold_r} are the constraints for fairness loss and adjustment overhead with threshold $\theta_1$ and $\theta_2$, where $\theta_1 \in [0,1]$, $\theta_2 \in [0,1]$. 
We can see that \textbf{P2}  is a typical MILP problem, which can be efficiently solved by standard MILP solves such as CPLEX. If there is no feasible solutions, Dorm would keep existing resource allocations until more running applications finish and release their resources.

\section{Numerical Results and Analysis} \label{sec:results}

In this section we investigate Dorm's performance using a testbed and popular distributed ML systems and applications.

\subsection{Experiments' Parameters and Configurations}
 
\subsubsection{Testbed Setup} The testbed contains 21 computation servers (1 DormMaster and 20 DormSlaves) and 2 storage servers connected by 10Gbps Ethernet. 
All training datasets are stored on the two storage servers. The DormMaster manages 240 CPU cores, 5 GPUs and 2.5TB RAM in total.

\subsubsection{Configurations} 

\renewcommand\arraystretch{1}
We use following thresholds for fairness loss and resource adjustment overhead in Dorm:
\vspace{-5pt}
\begin{table}[H]
\centering
\label{Tab: config}
\begin{tabular}{ | p{1cm} @{} p{1.4cm} @{} p{1.3cm}|}
\hline
Dorm-1  & $\; \; \theta_1 = 0.2$ &  $\; \; \theta_2 = 0.1$  \\ 
Dorm-2  & $\; \; \theta_1 = 0.1$ &  $\; \; \theta_2 = 0.2$ \\ 
Dorm-3  & $\; \; \theta_1 = 0.1$ &  $\; \; \theta_2 = 0.1$ \\ \hline
\end{tabular}
\end{table}
\vspace{-10pt}

\subsubsection{Workloads}  We generate an online workload based on the workload model of a production cluster in Sensetime. As shown in  Table \ref{Tab: job type}, the workload comprises  $50$ applications, which train $7$  ML models on public datasets.
We randomly submit them to Dorm  with a mean interval time of 20 minutes.

\renewcommand\arraystretch{1}
\begin{table} 
\centering
\resizebox{0.48 \textwidth}{!}{
\begin{threeparttable}
\caption{Synthetic workloads.}
\label{Tab: job type}
\begin{tabular}{@{} p{1.5cm} @{} p{1.5cm} @{} p{1.5cm} @{} p{1.3cm} @{}p{1cm}<{\centering} @{} p{0.7cm}<{\centering} @{}  p{0.7cm}<{\centering} @{} p{0.4cm}<{\centering}}
\hline
\textbf{Dependent System} & \textbf{Training Datasets} & \textbf{Trained Model$^*$} & \textbf{Resource Demand}$^\dagger$ &\textbf{Weight}  & \textbf{Max}  & \textbf{Min}  & \textbf{Num} \\ \hline
MxNet & Criteo-Log & LR &   2, 0, 8      & 1 & 32 & 1 & 20 \\
TensorFlow & MovieLens & MF  &  2, 0, 6   & 2 & 32 & 1 & 20 \\
MPI-Caffe & CIFAR-10 & CaffeNet  &  4, 0, 6  & 4 & 8 & 1 & 6 \\
MxNet & ImageNet & VGG-16  &  4, 1, 32   & 1 & 5 & 1 & 1 \\
TensorFlow & ImageNet & GoogLeNet  &  6, 1, 16  & 1 & 5 & 1 & 1 \\
Petuum & ImageNet & AlexNet   &  6, 1, 16   & 2 & 5 & 1 & 1 \\
MPI-Caffe & ImageNet & ResNet-50 
 &  4, 1, 32   & 4 & 5 & 1 & 1 \\ \hline
\end{tabular}
\begin{tablenotes}[normal,online]
\footnotesize
\item[$*$] LR: Logistic Regression; MF: Matrix Factorization.
\item[$\dagger$] Number of CPUs, number of GPUs and RAM size (GB).
\end{tablenotes}
\end{threeparttable}
}
\end{table}

\subsubsection{Baseline System} We use Swarm as the baseline system.  In the experiments,  Swarm would statically create 8, 8, 4, 2, 2, 2, 3 \emph{containers} for the $7$ types of applications in Table \ref{Tab: job type}.

\setlength{\minipagewidth}{0.24\textwidth}
\setlength{\figurewidthFour}{\minipagewidth}
\begin{figure} 
    \centering
    \begin{minipage}[t]{\minipagewidth}
    \begin{center}
    \includegraphics[width=\figurewidthFour]{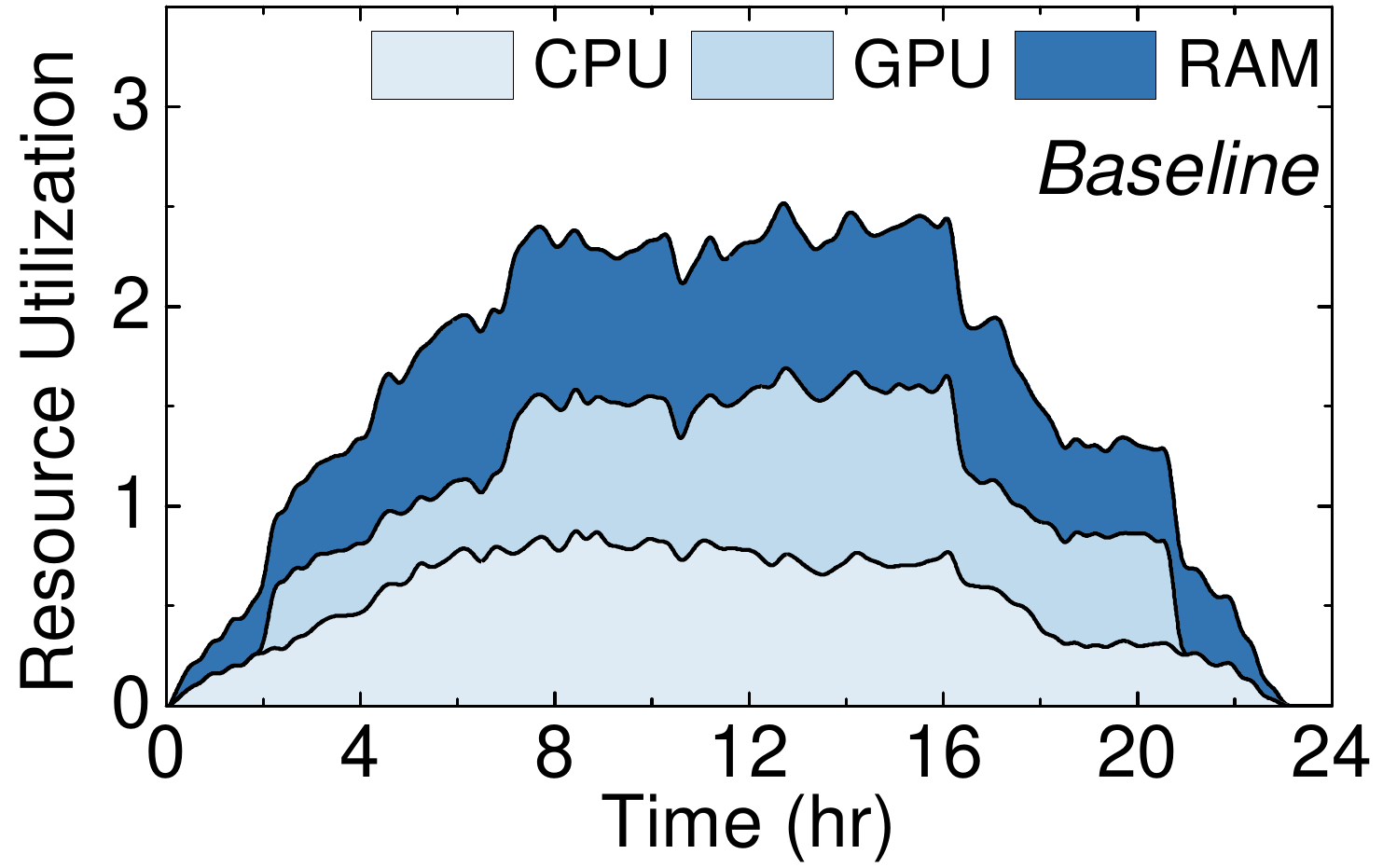}
        \vspace{0.5pt}
    \end{center}
    \end{minipage}
    \centering
    \begin{minipage}[t]{\minipagewidth}
    \begin{center}
    \includegraphics[width=\figurewidthFour]{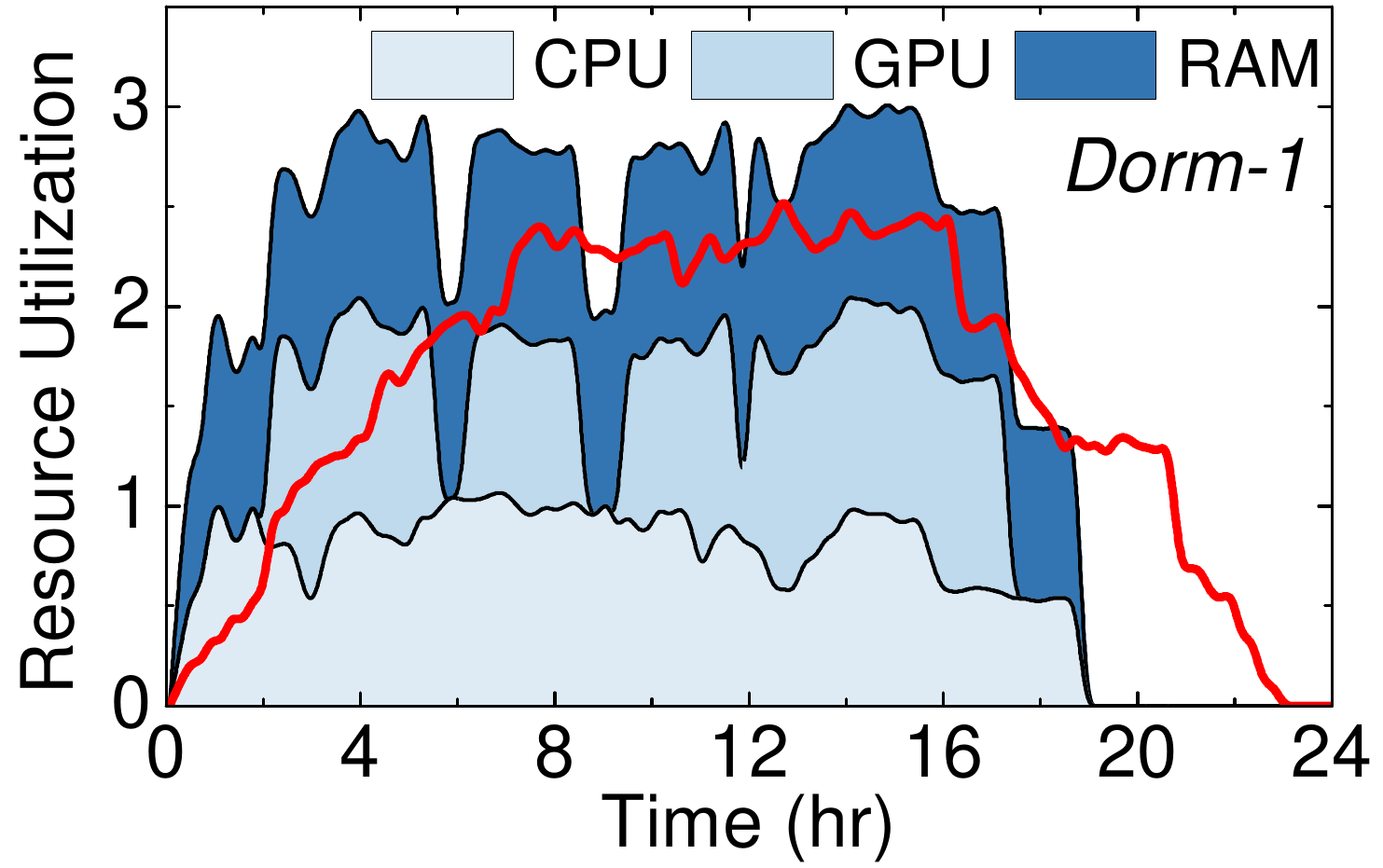}
    \end{center}
    \end{minipage}
    \centering
    \begin{minipage}[t]{\minipagewidth}
    \begin{center}
    \includegraphics[width=\figurewidthFour]{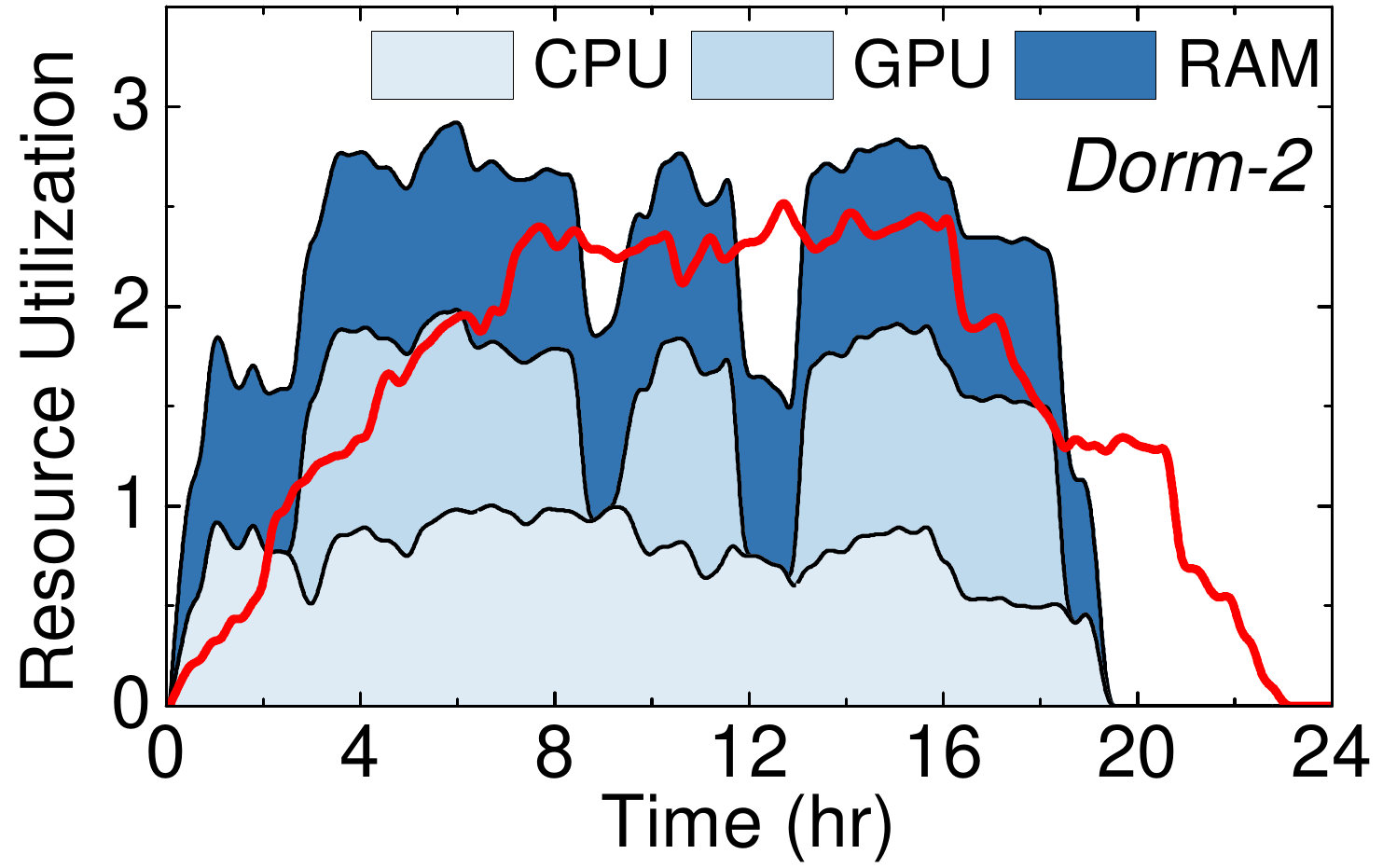}
    \end{center}
    \end{minipage}
    \centering
    \begin{minipage}[t]{\minipagewidth}
    \begin{center}
    \includegraphics[width=\figurewidthFour]{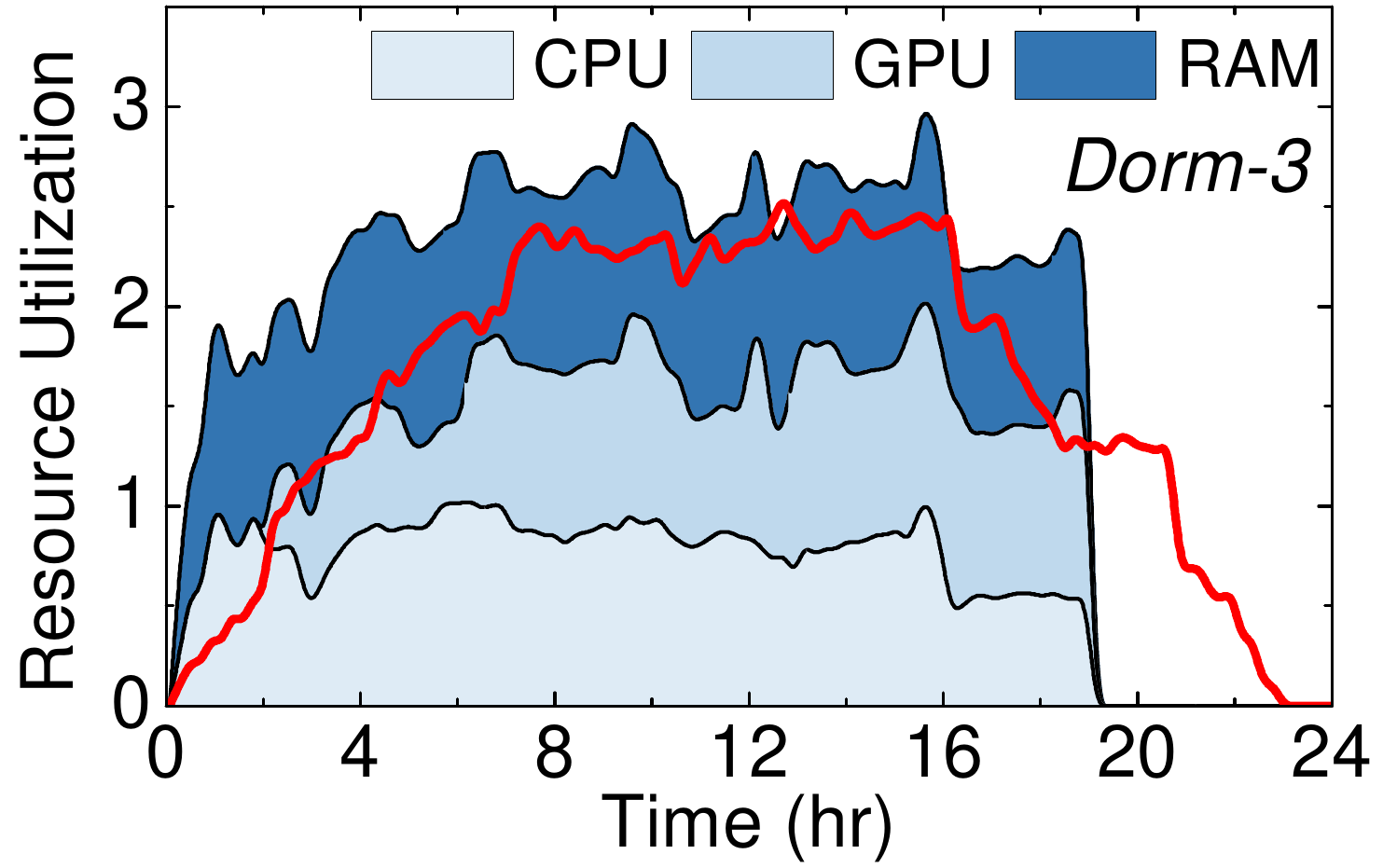}
    \end{center}
    \end{minipage}
    \centering
    \caption{Resource utilization of the testbed. The red line represents the overall resource utilization of the baseline system.}
\label{Fig: Resource_Util_Online}
\end{figure}

\subsection{Cluster Performance}

\subsubsection{Resource  Utilization}

Dorm could considerably improve the resource utilization, as shown in Figure \ref{Fig: Resource_Util_Online}. In the first $5$ hours, the baseline system has quite low resource utilization (which is up to 1.8), since it can only handle the first $15$ submitted applications based on their fixed resource requirements. In contrast, these applications could  dynamically scale up to take advantage of more resources on Dorm. As a result, compared to the baseline system, Dorm-1, Dorm-2 and Dorm-3 can increase the resource utilization by a factor of  2.55, 2.46 and 2.32 on average in the first 5 hours, respectively.

\subsubsection{Fairness Loss}

As shown in Figure \ref{Fig: Fairness_Loss_Online}, Dorm limits the fairness loss within a threshold, and can tolerate higher fairness loss with a larger $\theta_1$. 
Dorm-1 and  Dorm-3, which set $\theta_1$ to $0.2$ and $0.1$ with the same $\theta_2$, can limit the fairness loss within 1.5 and 0.6, respectively. 
Though Dorm-1 provides higher resource utilization than Dorm-3, its fairness loss is up to $1.78$ times higher than the baseline system. In contrast, Dorm-3 could reduce the fairness loss by a factor of 1.52 on average, compared to the baseline system. 

\setlength{\minipagewidth}{0.24\textwidth}
\setlength{\figurewidthFour}{\minipagewidth}
\begin{figure} 
    \centering
    \begin{minipage}[t]{\minipagewidth}
    \begin{center}
    \includegraphics[width=\figurewidthFour]{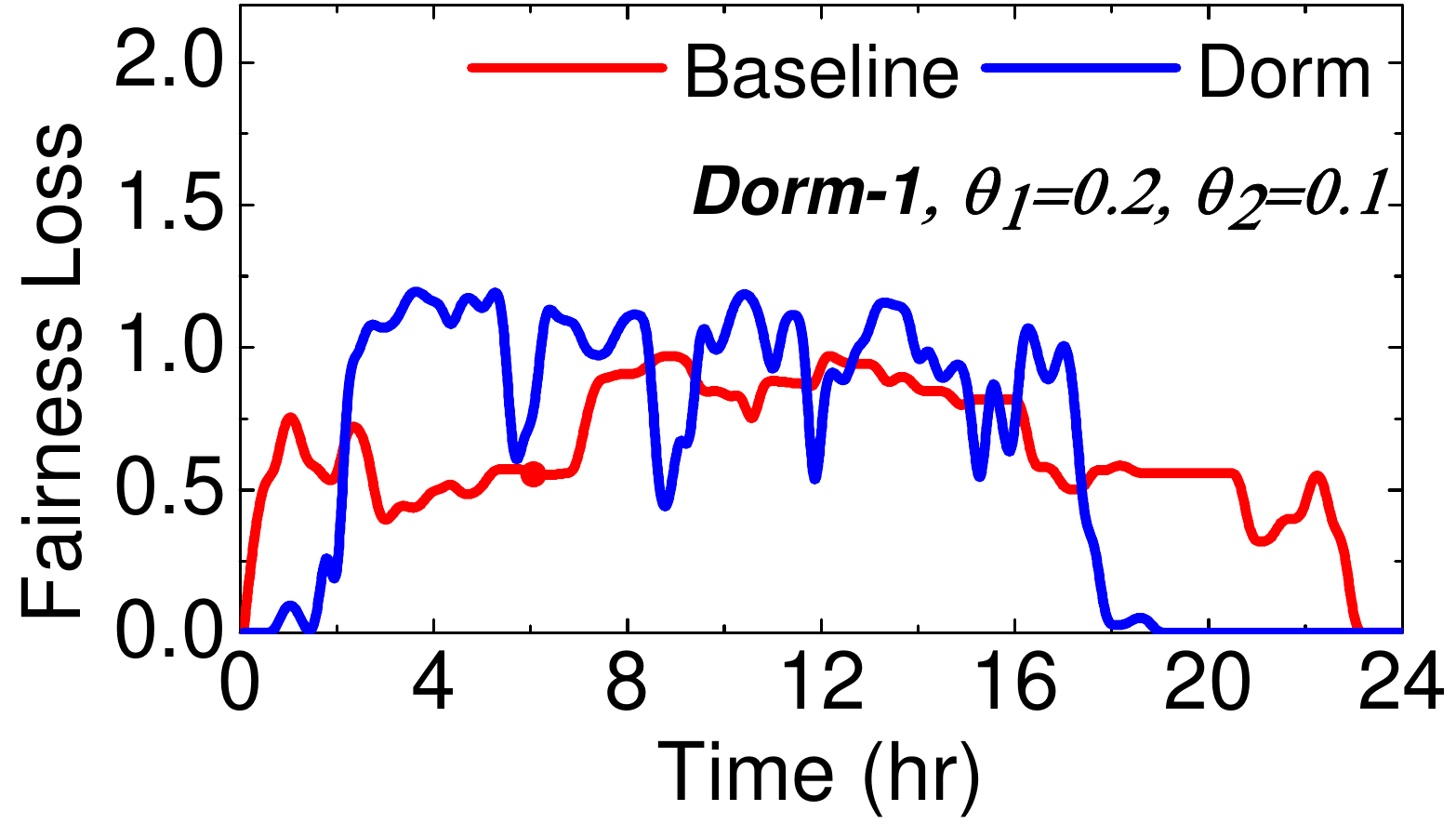}
    \end{center}
    \end{minipage}
    \centering
    \begin{minipage}[t]{\minipagewidth}
    \begin{center}
    \includegraphics[width=\figurewidthFour]{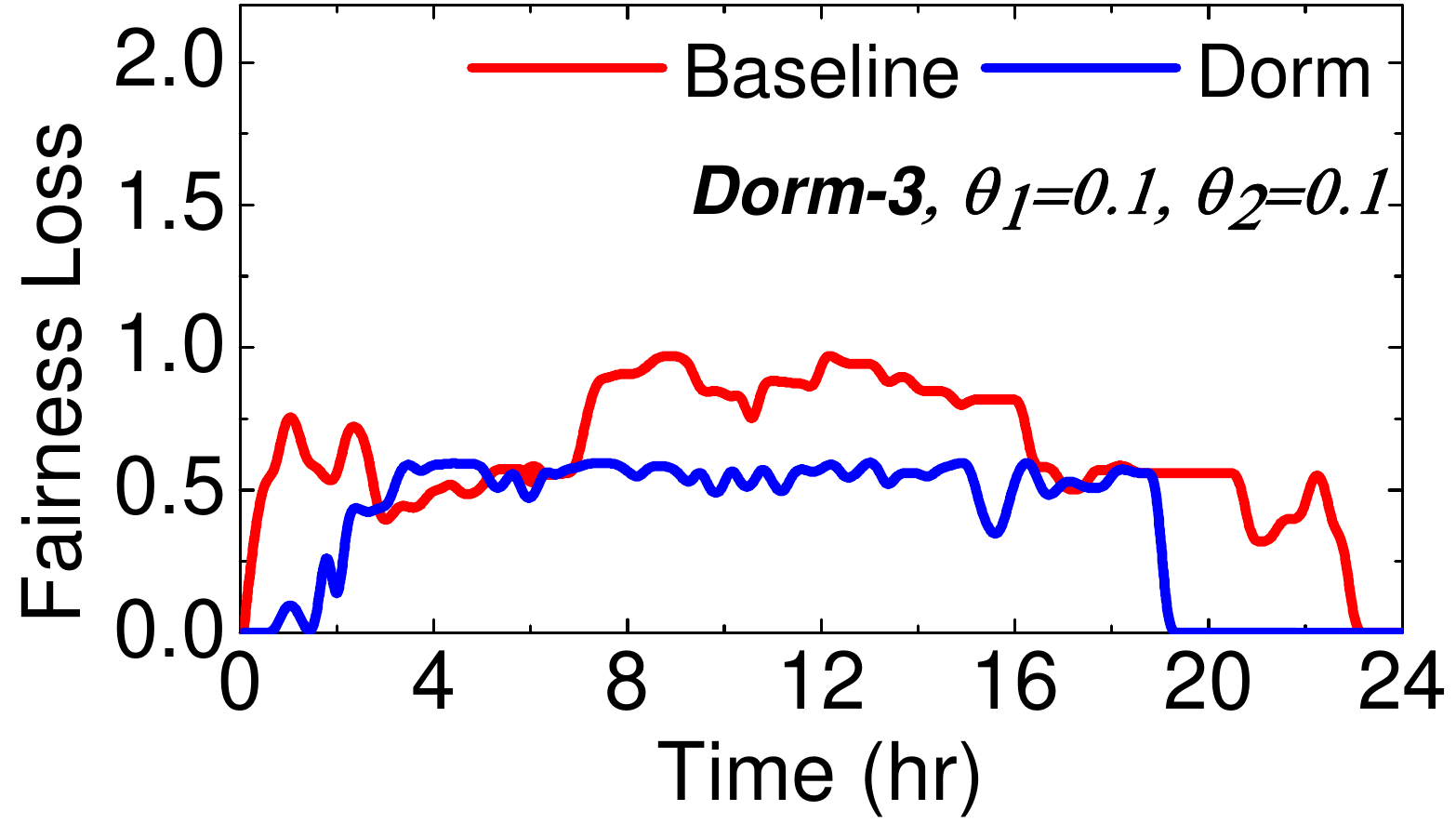}
    \end{center}
    \end{minipage}
    \centering
    \caption{Fairness loss of the testbed.}
\label{Fig: Fairness_Loss_Online}
\end{figure}

\setlength{\minipagewidth}{0.24\textwidth}
\setlength{\figurewidthFour}{\minipagewidth}
\begin{figure} 
    \centering
    \begin{minipage}[t]{\minipagewidth}
    \begin{center}
    \includegraphics[width=\figurewidthFour]{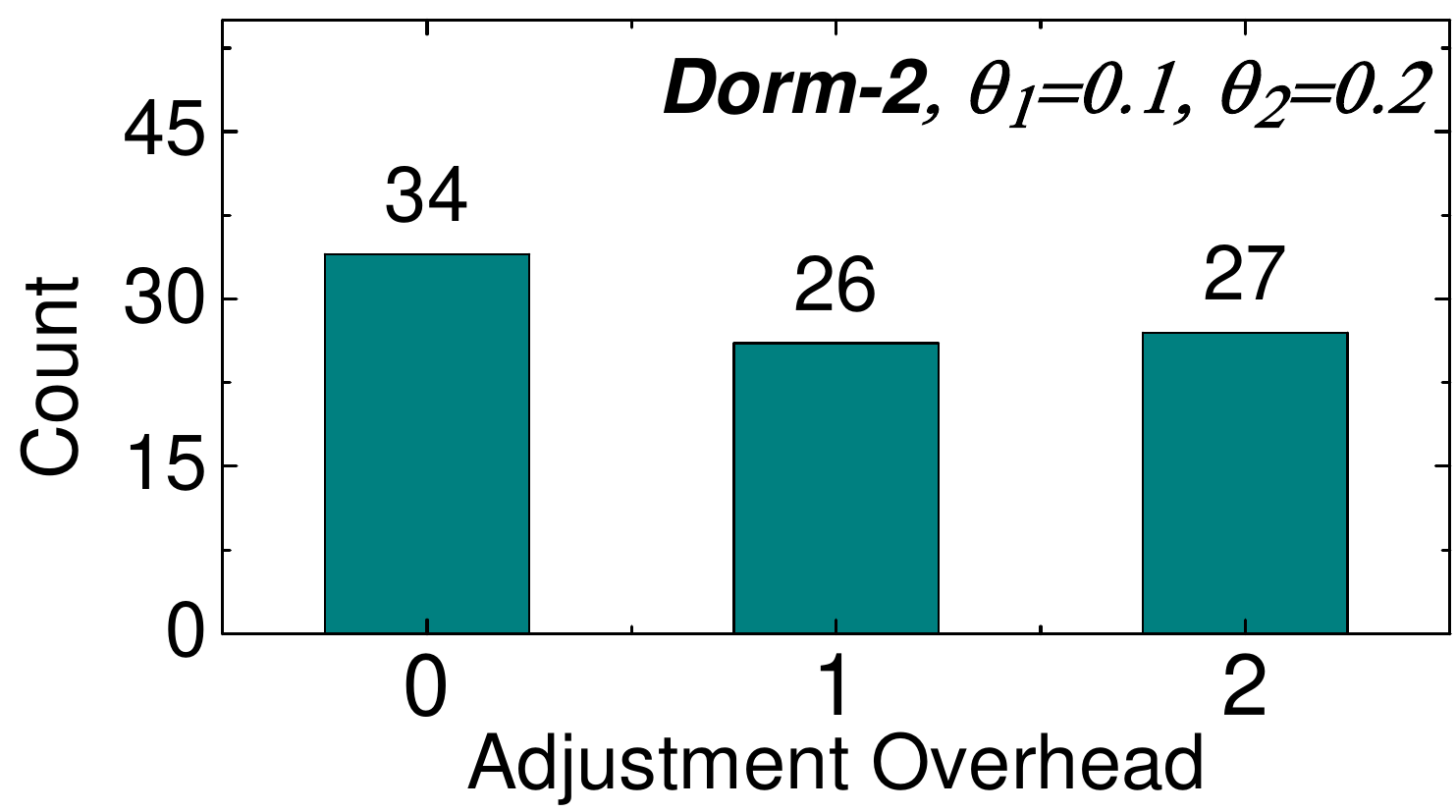}
    \end{center}
    \end{minipage}
    \centering
    \begin{minipage}[t]{\minipagewidth}
    \begin{center}
    \includegraphics[width=\figurewidthFour]{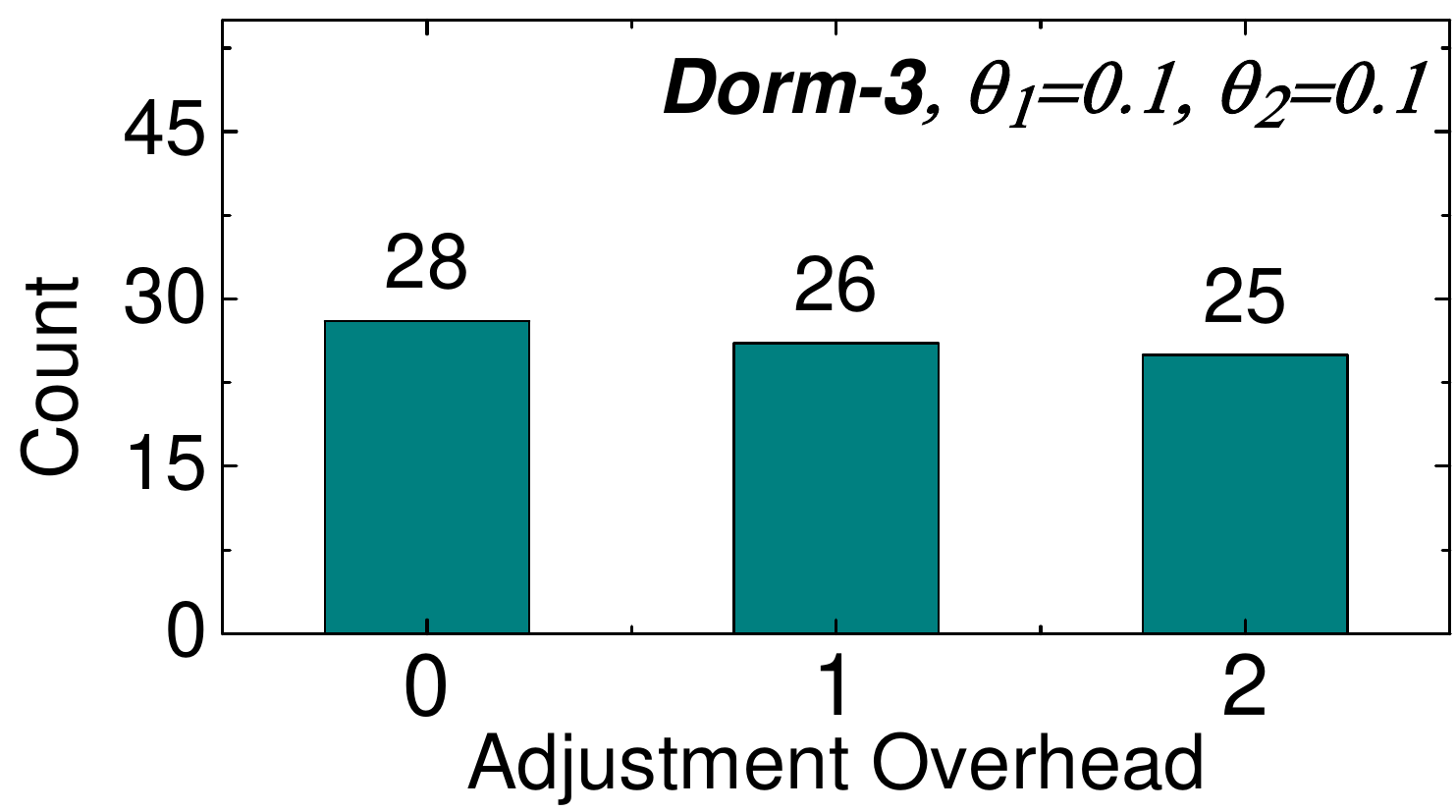}
    \end{center}
    \end{minipage}
    \centering
    \caption{Resource adjustment overhead of the testbed.}
\label{Fig: Adjust_Overhead_Online}
\end{figure}

\setlength{\minipagewidth}{0.232\textwidth}
\setlength{\figurewidthFour}{\minipagewidth}
\begin{figure} 
    \centering
    \begin{minipage}[t]{\minipagewidth}
    \begin{center}
    \includegraphics[width=\figurewidthFour]{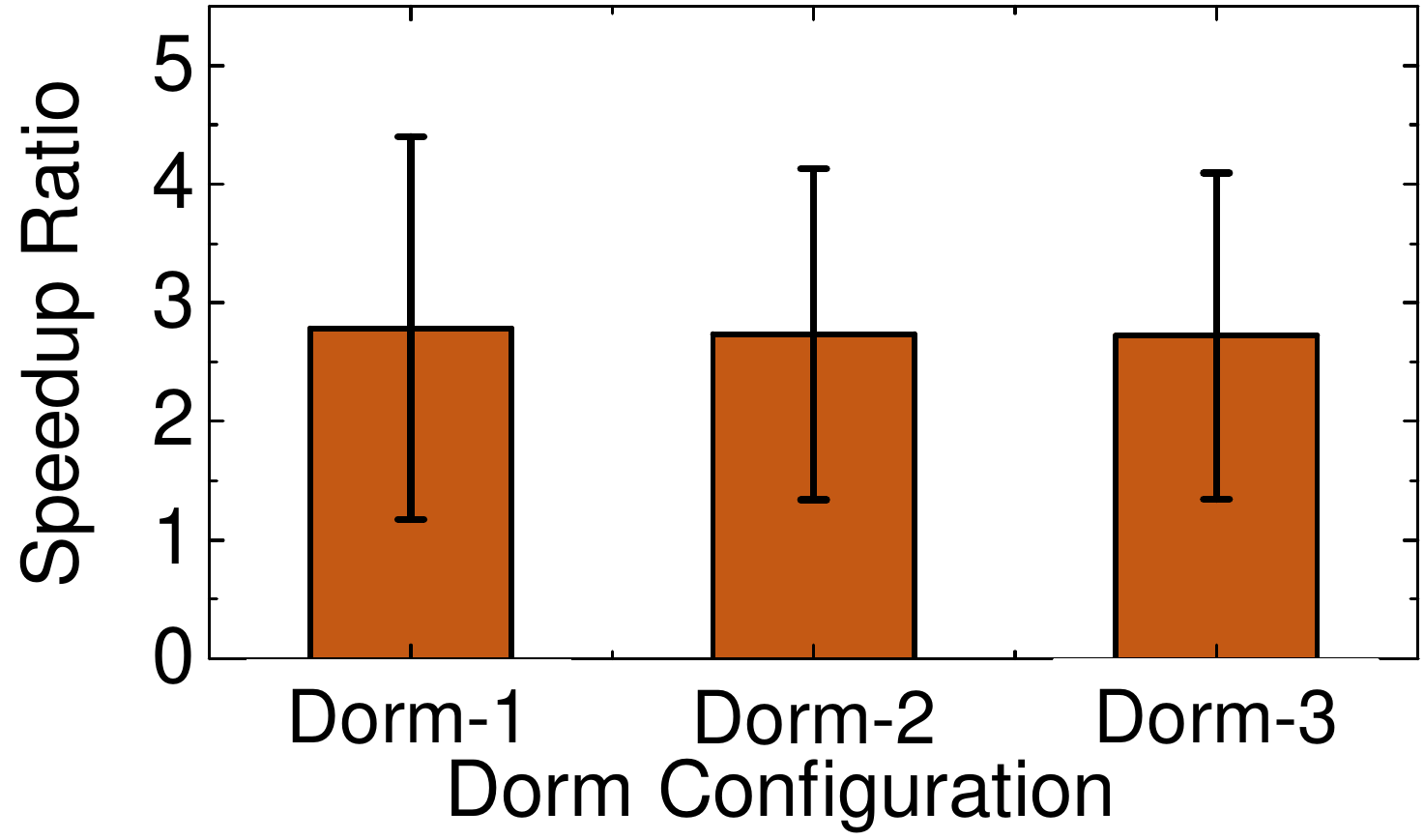}
    \subcaption{(a) Application Speedup Ratio}
    \end{center}
    \end{minipage}
    \centering   
    \hspace{1 mm}  
    \begin{minipage}[t]{\minipagewidth}
    \begin{center}
    \includegraphics[width=\figurewidthFour]{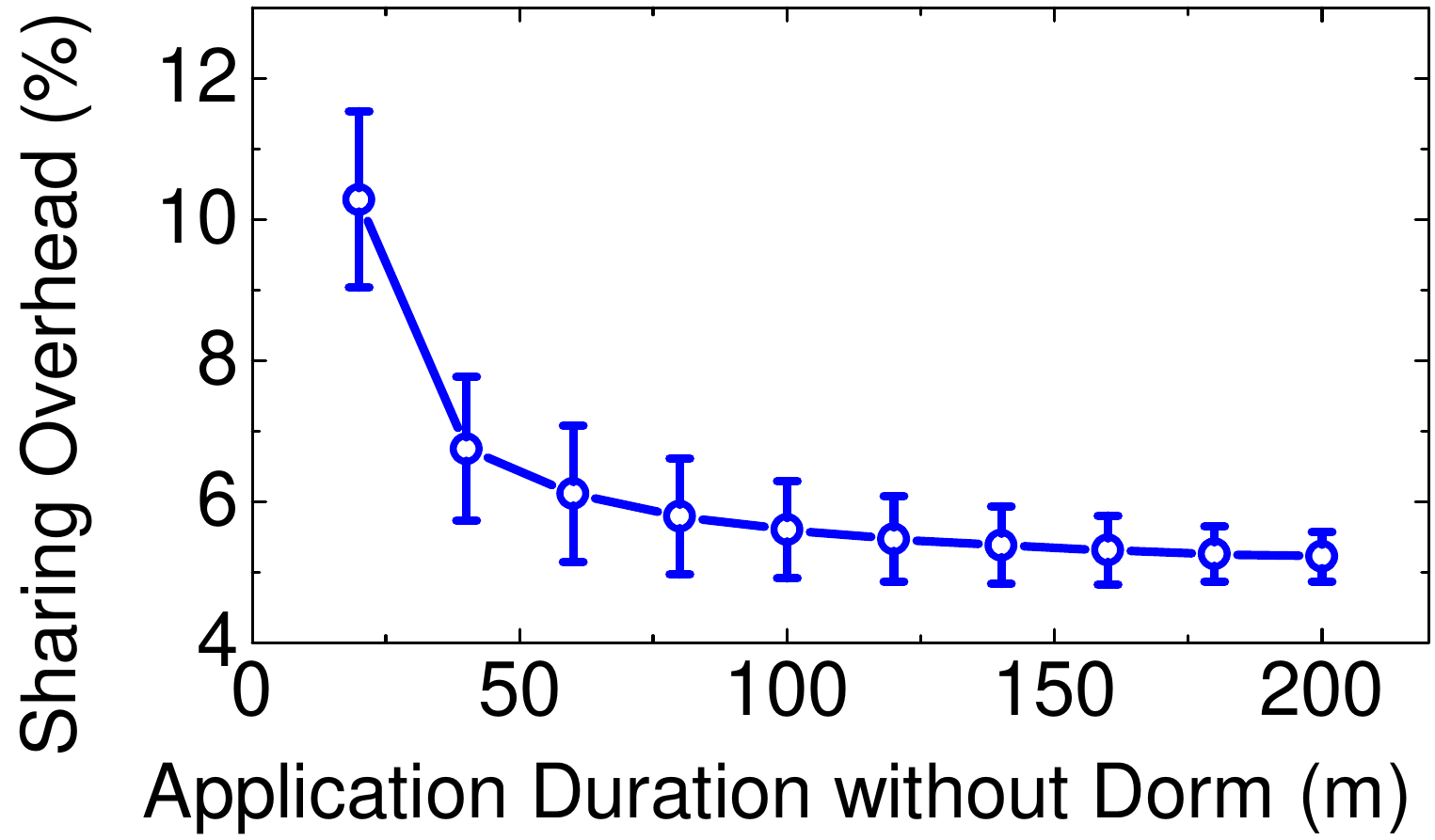}
    \subcaption{(b) Sharing Overhead}
    \end{center}
    \end{minipage}
    \centering  
    \vspace{-10pt}      
    \caption{Application speedup ratio and Dorm's sharing overhead.}
\label{Fig: scheduling_overhead}
\end{figure}

\subsubsection{Resource Adjustment Overhead}

Figure \ref{Fig: Adjust_Overhead_Online} shows that  Dorm can limit the resource adjustment overhead within a threshold, and can tolerate higher resource adjustment overhead with a larger $\theta_2$.  Dorm-2 and Dorm-3, which set $\theta_2$ to $0.2$ and $0.1$ with the same $\theta_1$, would kill and resume 2 applications at most per resource adjustment operation, and  affect 80 and 76 applications in total in 24 hours, respectively.

\subsubsection{Speedup Ratio}

Distributed ML applications running on Dorm consistently perform better than those running on the baseline system. Figure \ref{Fig: scheduling_overhead}(a) shows that Dorm-1, Dorm-2 and Dorm-3  can speed up distributed ML applications by a factor of 2.79, 2.73 and 2.72 on average, respectively. 

\subsubsection{Sharing Overhead}
To measure Dorm's sharing overhead, we compare applications'  performances in two cases.
First, we set up  a dedicated MxNet cluster on 10 worker nodes (each node has 4 CPUs and 16GB RAM), and run a set of applications on it. We then submit  same applications to Dorm  with the same amount of resources (i.e., $n_{max}=n_{min}=10$, and each \emph{container} has 4 CPUs and 16GB RAM).  During the application running time, our tested MxNet-based applications are randomly killed and resumed 2 times on Dorm.

Dorm's sharing overhead is not significant for distributed ML applications.
As shown in Figure \ref{Fig: scheduling_overhead}(b), when the application duration is longer than 3 hours, Dorm would roughly increase the application duration by a factor of 1.05 (i.e., the sharing overhead of Dorm is about $5\%$).  Compared to the performance gain, Dorm's sharing overhead is acceptable.

\section{Summary} \label{sec: summary}

We propose a novel cluster management system named Dorm to efficiently and fairly share a single cluster among  distributed ML applications with low sharing overhead.  To achieve this goal, Dorm employs a dynamically-partitioned sharing model and an utilization-fairness optimizer. We have implemented Dorm and enabled it to work with Petuum, MxNet, TensorFlow and MPI-Caffe. In the future, we plan to integrate it with more distributed ML systems and applications.

\balance

\bibliographystyle{IEEEtran}
\bibliography{main}

\begin{thebibliography}{10}
\providecommand{\url}[1]{#1}
\csname url@samestyle\endcsname
\providecommand{\newblock}{\relax}
\providecommand{\bibinfo}[2]{#2}
\providecommand{\BIBentrySTDinterwordspacing}{\spaceskip=0pt\relax}
\providecommand{\BIBentryALTinterwordstretchfactor}{4}
\providecommand{\BIBentryALTinterwordspacing}{\spaceskip=\fontdimen2\font plus
\BIBentryALTinterwordstretchfactor\fontdimen3\font minus
  \fontdimen4\font\relax}
\providecommand{\BIBforeignlanguage}[2]{{%
\expandafter\ifx\csname l@#1\endcsname\relax
\typeout{** WARNING: IEEEtran.bst: No hyphenation pattern has been}%
\typeout{** loaded for the language `#1'. Using the pattern for}%
\typeout{** the default language instead.}%
\else
\language=\csname l@#1\endcsname
\fi
#2}}
\providecommand{\BIBdecl}{\relax}
\BIBdecl

\bibitem{chen2015mxnet}
T.~Chen, M.~Li, Y.~Li, M.~Lin, N.~Wang, M.~Wang, T.~Xiao, B.~Xu, C.~Zhang, and
  Z.~Zhang, ``Mxnet: A flexible and efficient machine learning library for
  heterogeneous distributed systems,'' \emph{arXiv preprint arXiv:1512.01274},
  2015.

\bibitem{jia2014caffe}
Y.~Jia, E.~Shelhamer, J.~Donahue, S.~Karayev, J.~Long, R.~Girshick,
  S.~Guadarrama, and T.~Darrell, ``Caffe: Convolutional architecture for fast
  feature embedding,'' in \emph{ACM MM 14}.\hskip 1em plus 0.5em minus
  0.4em\relax ACM, 2014, pp. 675--678.

\bibitem{abadi2016tensorflow}
M.~Abadi, P.~Barham, J.~Chen, Z.~Chen, A.~Davis, J.~Dean, M.~Devin,
  S.~Ghemawat, G.~Irving, M.~Isard, M.~Kudlur, J.~Levenberg, R.~Monga,
  S.~Moore, D.~G. Murray, B.~Steiner, P.~Tucker, V.~Vasudevan, P.~Warden,
  M.~Wicke, Y.~Yu, and X.~Zheng, ``Tensorflow: A system for large-scale machine
  learning,'' in \emph{OSDI 16}.\hskip 1em plus 0.5em minus 0.4em\relax USENIX,
  Nov. 2016, pp. 265--283.

\bibitem{xing2015petuum}
E.~P. Xing, Q.~Ho, W.~Dai, J.~K. Kim, J.~Wei, S.~Lee, X.~Zheng, P.~Xie,
  A.~Kumar, and Y.~Yu, ``Petuum: a new platform for distributed machine
  learning on big data,'' \emph{Big Data, IEEE Transactions on}, vol.~1, no.~2,
  pp. 49--67, 2015.

\bibitem{hu2014toward}
H.~Hu, Y.~Wen, T.-S. Chua, and X.~Li, ``Toward scalable systems for big data
  analytics: A technology tutorial,'' \emph{IEEE Access}, vol.~2, pp. 652--687,
  2014.

\bibitem{hindman2011mesos}
B.~Hindman, A.~Konwinski, M.~Zaharia, A.~Ghodsi, A.~D. Joseph, R.~H. Katz,
  S.~Shenker, and I.~Stoica, ``Mesos: A platform for fine-grained resource
  sharing in the data center.'' in \emph{NSDI 11}, 2011, pp. 22--22.

\bibitem{delimitrou2014quasar}
C.~Delimitrou and C.~Kozyrakis, ``Quasar: Resource-efficient and qos-aware
  cluster management,'' \emph{ACM SIGPLAN Notices}, vol.~49, no.~4, pp.
  127--144, 2014.

\bibitem{sefraoui2012openstack}
O.~Sefraoui, M.~Aissaoui, and M.~Eleuldj, ``Openstack: Toward an open-source
  solution for cloud computing,'' \emph{International Journal of Computer
  Applications}, vol.~55, no.~3, 2012.

\bibitem{vavilapalli2013apache}
V.~K. Vavilapalli, A.~C. Murthy, C.~Douglas, S.~Agarwal, M.~Konar, R.~Evans,
  T.~Graves, J.~Lowe, H.~Shah, S.~Seth \emph{et~al.}, ``Apache {Hadoop Yarn}:
  Yet another resource negotiator,'' in \emph{SOCC 13}.\hskip 1em plus 0.5em
  minus 0.4em\relax ACM, 2013, p.~5.

\bibitem{yin2013cloud3dview}
J.~Yin, P.~Sun, Y.~Wen, H.~Gong, M.~Liu, X.~Li, H.~You, J.~Gao, and C.~Lin,
  ``Cloud3dview: an interactive tool for cloud data center operations,'' in
  \emph{ACM SIGCOMM Computer Communication Review}, vol.~43, no.~4.\hskip 1em
  plus 0.5em minus 0.4em\relax ACM, 2013, pp. 499--500.

\bibitem{jin2013empirical}
Y.~Jin, Y.~Wen, Q.~Chen, and Z.~Zhu, ``An empirical investigation of the impact
  of server virtualization on energy efficiency for green data center,''
  \emph{The Computer Journal}, vol.~56, no.~8, pp. 977--990, 2013.

\bibitem{verma2015large}
A.~Verma, L.~Pedrosa, M.~Korupolu, D.~Oppenheimer, E.~Tune, and J.~Wilkes,
  ``Large-scale cluster management at google with borg,'' in \emph{EuroSys
  15}.\hskip 1em plus 0.5em minus 0.4em\relax ACM, 2015, p.~18.

\bibitem{schwarzkopf2013omega}
M.~Schwarzkopf, A.~Konwinski, M.~Abd-El-Malek, and J.~Wilkes, ``Omega:
  Flexible, scalable schedulers for large compute clusters,'' in \emph{EuroSys
  13}.\hskip 1em plus 0.5em minus 0.4em\relax ACM, 2013, pp. 351--364.

\bibitem{boutin2014apollo}
E.~Boutin, J.~Ekanayake, W.~Lin, B.~Shi, J.~Zhou, Z.~Qian, M.~Wu, and L.~Zhou,
  ``Apollo: Scalable and coordinated scheduling for cloud-scale computing,'' in
  \emph{OSDI 14}, 2014, pp. 285--300.

\bibitem{ousterhout2013sparrow}
K.~Ousterhout, P.~Wendell, M.~Zaharia, and I.~Stoica, ``Sparrow: Distributed,
  low latency scheduling,'' in \emph{SOSP 13}.\hskip 1em plus 0.5em minus
  0.4em\relax ACM, 2013, pp. 69--84.

\bibitem{delgado2015hawk}
P.~Delgado, F.~Dinu, A.-M. Kermarrec, and W.~Zwaenepoel, ``Hawk: Hybrid
  datacenter scheduling,'' in \emph{USENIX ATC 15}, 2015, pp. 499--510.

\bibitem{karanasos2015mercury}
K.~Karanasos, S.~Rao, C.~Curino, C.~Douglas, K.~Chaliparambil, G.~M. Fumarola,
  S.~Heddaya, R.~Ramakrishnan, and S.~Sakalanaga, ``Mercury: Hybrid centralized
  and distributed scheduling in large shared clusters,'' in \emph{USENIX ATC
  15}, 2015, pp. 485--497.

\bibitem{ghodsi2011dominant}
A.~Ghodsi, M.~Zaharia, B.~Hindman, A.~Konwinski, S.~Shenker, and I.~Stoica,
  ``Dominant resource fairness: Fair allocation of multiple resource types.''
  in \emph{NSDI 13}, vol.~11, 2011, pp. 24--24.

\end{thebibliography}


\end{document}